\documentclass{elsart}

\usepackage{epsfig}

\usepackage{amssymb}
\journal{Journal of Computational physics}

\begin{document}

\begin{frontmatter}



 \title{\thanksref{label1}Constrained Molecular Dynamics II: an N-body 
 approach to nuclear systems }
\thanks[label1]{
We thank Mr. F.Ferrera, G.Del Tevere (L.N.S.) and C.Rocca (I.N.F.N. 
Sezione di Catania) 
belonging to the computing support staff
for the kind assistance. Finally we thank Dr. J.Winkler for the careful revision of the English grammar in the text.}

\author[South]{M.Papa\corauthref{cor}}
\corauth[cor]{corresponding author.}
\ead{Massimo.Papa@ct.infn.it}
\author[Oxford]{G.Giuliani}
\author[Oxford1]{A.Bonasera}
\address[South]{Istituto Nazionale di Fisica Nucleare Sezione di Catania
V.S.Sofia 64 95133 Catania (Italy)}
\address[Oxford]{Universita' degli studi di Catania -
 dipartimento di Fisica e Astronomia V.S.Sofia 64 95133 Catania (Italy)}
\address[Oxford1]{Istituto Nazionale di Fisica Nucleare- Laboratorio
Nazionale del SUd V.S.Sofia 44 95133 Catania (Italy)}


\begin{abstract}
In this work, we illustrate the basic development of the constrained molecular dynamics applied to the N-body problem in nuclear physics. 
The heavy computational tasks related to quantum effects, to the Fermionic
nature of the system  have been resolved out by defining a set of transformations based on the concept of impulsive forces. 
In particular, in the implemented version II of the Constrained Molecular Dynamics model the problem related to the non conservation of the total angular momentum has been solved.
This problem affects other semi-classical microscopic approaches due to the "hard core"  repulsive interaction and, more generally, to the usage of random forces. 
The effect of the restored conservation law  on the fusion cross-section for the $^{40}Ca+^{40}Ca$ system is also  briefly discussed.
\end{abstract}


\end{frontmatter}

\section{Introduction}
In the present work we will try to illustrate, with concrete examples, the strategy used to solve the computational task related to the quantum N-body problem. 
In particular, we will propose a solution based on the concept of the constrained dynamics and we will focus on the treatment of the many-body problem  for nuclear-systems. 
In the following, we will give an introduction by briefly explaining  the main features of such complex systems and some of the crucial problems that should be solved to obtain a reliable description.

Nuclear systems are under different aspects unique in nature. 
In such systems can coexist different aspects of the physical laws whose interplay, in general, cannot be neglected as instead can happen for other complex systems. 
Nuclei are complex systems made of an ensemble of N nucleons whose number can vary from some units up to several tens of protons and neutrons.
Therefore, they are in general not so simple that the particles follow the behaviors suggested by the few body models, and not so large and so complex that they follow the features suggested by the conventional statistical mechanics or by the hydrodynamics models \cite{comd1}.  

Nuclei are sub-atomic objects and therefore in a wide range of energies their behavior is strongly dominated by quantum effects. 
Their constituents are strong interacting fermions through the short-range nuclear interaction, but they interact also through the long-range proton-proton Coulomb repulsion. 
Still, the force, which is responsible for their stability and  for the evolution of enormous objects like stars, is not well known, especially at high energy when pieces of hot and compressed nuclear matter can be produced in extreme conditions. 
Up to now, these forces are not derived from a general theoretical frame.
This obviously on one hand makes still necessary and intriguing the study of such systems, but on the other one adds an uncertainty to the structure of the related many  body Hamiltonian.

The models can aim to describe such complex situation, especially around and above the Fermi energy, can be only N-body approaches (N being the number of nucleons belonging to the system). 
As happens for all the N- body approaches, two main roads can be followed. 

One of these  can be derived from the elegant hierarchy of kinetic equations developed in the early twentieth century.
This approach deals with the theoretical N-body  distribution 
function $F^{N}(1,2......,..N,t)$ which gives the probability that N nucleons  occupy the region of the phase space ${\bf r}_{1},{\bf p}_{1},{\bf r}_{2},{\bf p}_{2}...{\bf r}_{N},{\bf p}_{N}$ at the time t (with bold symbols we will indicate vector quantities).
Up to now, the kinetic equations that have been simulated in real three-dimensional problems are the Boltzmann equation, which represents the first equation of the hierarchy in the classical case, and the BUU and BNV\cite{buu,bon} corresponding approaches including some quantum corrections. 
Nevertheless these equations deal with $F^{1}(1,t)$ representing the probability that one of the N-nucleons has coordinates ${\bf r}_{1},{\bf p}_{1}$ at time t, independently from the position of the other, N-1 particles.   
It therefore gives quite limited information with respect to $F^{N}$. 
By definition  $F^{1}$ can give predictions on the averages of one-body quantities (mean field-approach) but it cannot describe  N-body correlations that are decisive in studying, for example, deviations from the averages or phenomena like the cluster formation processes.

Another much more pragmatic road to follow, which can take into account N-body correlations, gives arise to the so-called molecular dynamics approaches. The above mentioned quantity $F^{N}$, in the semi classical approximation scheme, has to be understood as an ensemble averaged probability computed on many replications (in the theoretical limit an infinite number) of the same systems characterized by a total energy E, total angular momentum $\bf{L}$, total mass N, total charge Z  or other quantities which define the initial macroscopic state. It is not difficult to think that for an N-body system many microscopic configurations can describe the same total system in the same macroscopic state. This happens from a classical point of view and at high excitation energy also from a quantal point of view. 
Therefore, in this situation, in the classical case for example, if we  choose a kind of nucleon-nucleon interactions we can solve  with the computer the Newtonian equations of motion and follow in time the single realization or event. 
Each of these events will have a proper history and if their number is enough large we can aim to have a good estimation of the true $F^{1}$, $F^{2}$,...$F^{N}$.
The concept between model and simulation with the computer therefore becomes closely connected. This connection becomes also stronger when particular hypothesis are necessary to be done to make the problems solvable numerically in a reasonable time.
In fact to compare results with experiments we need to "compute" as many events as possible. In such way we can neglect the errors related to the statistics of simulations, and this should be done as fast as possible to make practicable our investigations.
In the following, we will illustrate a recent development of the molecular dynamics approach, which is based on the concept of constraint. 
In the Constrained Molecular Dynamics approach (CoMD) \cite{comd,comd1}, we force our equations of motion to satisfy two quite fundamental conditions that have been undertaken in the past molecular dynamics models: 

The first condition is directly related to the quantum nature of the system.
We force the average occupation numbers $\overline{f_{i}}$ to satisfy the condition $\overline{f_{i}}\leq 1$ at each time step, according to the Pauli principle. 
Therefore, in the model-simulation our system can reflect the properties of a quantum liquid drop. 
This crucial point \cite{comd}, and the main ingredients of the model will be illustrated from section 2 to section 5.

Another problem is related to the strong repulsive character of nuclear interaction at short distance and more generally to the usage of stochastic process in dynamical models. 
The strategy used to overcome the related numerical problem in most of the transport simulation codes  gives arise to a dynamical evolution that does not preserve the total angular momentum.
In section 6, we will discuss in some detail this point and we will present a numerical algorithm able to restore this fundamental conservation rule. 
In section 7 the related effects on the fusion cross-section and binary process yields induced on the $^{40}Ca+^{40}Ca$ system will be also illustrated.

\section{Molecular Dynamics models in nuclear physics}

As observed in the previous section, one of the crucial problems in the molecular dynamics approach for fermionic system is to reproduce the correlations induced by the Pauli principle. 
To understand better the way in which we have introduced such correlations we briefly recall the main ingredients of the semi-classical molecular dynamics approaches in nuclear physics \cite{qmd,comd}.
In this approach, each nucleon belonging to one possible replication of our N-body system is represented by a Gaussian distribution function \cite{qmd,comd}.
\begin{eqnarray}
f_i({\bf r},{\bf p}) &=&  {1 \over (2\pi\sigma_{r}\sigma_{p})^{3}} \cdot
              \exp\left[-{({\bf r}-\langle{\bf r}\rangle_i)^2\over 
2{\sigma_r}^2}
            -{2{\sigma_r}^2({\bf p}-\langle{\bf p}\rangle_i)^2\over
2{\sigma_p}^2}\right].
\end{eqnarray}
where ${\sigma_r}$ and ${\sigma_p}$ represent the dispersions in the configuration and in the momentum space, respectively.
Starting from the above distribution, we can define the occupation density
in phase space as \cite{comd}:
\begin{eqnarray}
\overline{f}_{i} &\equiv&
 \sum_j \delta(\tau_{i}-\tau_{j})
  \delta(s_{i}-s_{j})\int_{h^{3}}f_{j}({\bf r}_{j}, {\bf p}_{j};
  \langle {\bf r}_i\rangle,\langle {\bf p}_i\rangle)\;d^3r_j\;d^3p_j
\end{eqnarray}
The coordinates $s_{i}$ and $\tau_{i}$ represent the nucleon spin and isospin (nuclear charge) projection quantum number.
The integral is performed in a hypercube of volume $h^{3}$ in the phase-space centered around the point $(\langle {\bf r}_i\rangle,\langle {\bf p}_i\rangle)$ with size $\sqrt{ {2\pi\hbar \over \sigma_{r}\sigma_{p}} }\sigma_{r}$ and
$\sqrt{ {2\pi\hbar \over \sigma_{r}\sigma_{p}} }\sigma_{p}$ in the  
$r$ and $p$ spaces respectively.
The N-body distribution function for each replication $k$ in phase space is expressed as the direct product:
\vskip 1pt
\begin{eqnarray}
f^{N,k}(1,2,..N,t) &=&  f_1({\bf r_{1}},{\bf p_{1}}) 
\cdot f_2({\bf r_{1}},{\bf p_{2}})
....f_N({\bf r_{N}},{\bf p_{N}})
\end{eqnarray}
Each realization is identified by the N-uplet of coordinates in phase space that determine the positions of the N Gaussian. 
The link between the previous expression and the N-body distribution function is clearly obtained through the ensemble average procedure on a very large number of realizations M of the same system \cite{comd1}:
\begin{eqnarray}
F^{N}(1,2......,..N,t)=\lim_{M \rightarrow 
\infty}\sum_{k=1}^{M}{f^{N,k} \over M} 
\end{eqnarray}
This expression can be made symmetric for N identical particles by averaging over the permutations P related to the N-uplet of coordinates in phase space.
For each replication the equations of motion for the wave packet centers $\langle{\bf r}\rangle_{i}$, $\langle{\bf p}\rangle_{i}$ related to $f_{i}({\bf r},{\bf p})$ are derived using the time-dependent variational principle
\cite{qmd}, which gives:
\begin{equation}
\dot{\langle{\bf r}\rangle}_i =   \frac{\partial H}{\partial
\langle{\bf p}\rangle_i},
\;\;\;\
\dot{\langle{\bf p}\rangle}_i = - \frac{\partial H}{\partial
\langle{\bf r}\rangle_i}+{\bf C}_{i}.
\label{eq4}
\end{equation}

The Hamiltonian $H$ consists of the kinetic energy and the two-body effective interaction:
\begin{equation}
H=\sum_i {\langle{\bf p}\rangle_i^2\over 2m} +{1\over 2}\sum_{i,j\neq i}
V_{ij}+3{\sigma_p^2 \over 2m}\\
\end{equation}
where $V_{ij}$ represents the two body interaction between particles $i$ and $j$. It is obtained by folding the effective nucleon-nucleon interaction with the nucleon distribution. More details in the case of a Skyrme interaction can be found, for example, in ref. \cite{comd,qmd}.

The quantity ${\bf C}_{i}$ represents a random impulsive force which simulates by a scattering process, the effect of the nucleon-nucleon repulsive short-range interaction. 
At each time step, if two nucleons are within an interaction
radius $R_{int}$ they can scatter with a probability determined by the related cross section $\sigma_{nn}$ and the Pauli blocking attenuation factor $P_{b}$ evaluated for the final state \cite{buu,bon,qmd,comd}. A correct evaluation of $P_{b}$ can be obtained only if the numbers ${\overline{f}_{i}}$ defined in eq.(2) are at each time less then or equal to 1.
\vskip 1pt
It is easy to see that the illustrated scheme is essentially classical. 
There are no reasons for the Pauli condition to be satisfied. 
As it was shown in ref.\cite{comd} after some tens of fm/c large deviations from the Pauli requirement (even if the nuclei are correctly initialized) affect about 50$\%$ of the total system.

To solve the above problem, more complex approaches have been developed \cite{fmd,amd}.
In these cases the trial wave function, which represents the many body system, is given by the anti-symmetric product of wave packets.
This rigorous way to solve the problem, introduce a great amount of complexity. 
While the structure of the equations of motion (5), with two-body interactions, implies a $N^{2}$ dimensionality, the use of the anti-symmetric structure for the total wave function increases the dimensionality to $N^{4}$. Normally therefore some approximations are necessary to work out calculations like  the so-called "physical coordinate approximation" \cite{amd}.
It appears clear that, as an example, for a nucleus with 200  nucleons,  an exact anti-symmetric scheme should naturally produce a computation speed $4\cdot 10^{4}$ slower with respect to a scheme based on the direct product. 
This is the reason why in our CoMD approach we solve the Pauli principle requirement by keeping the $N^{2}$ dimensionality.


We want to conclude this section by observing that the lack of the Pauli principle has also catastrophic consequences on the structure of the "ground states" (g.s.) of nuclei. 
It is easily understood that the Hamiltonian given in eq.(7) admits a minimum energy  state for which the total effective kinetic energy (first term in eq.(6)) related to the centers of the wave packet is zero.
In this case the total energy is given by the potential energy plus the bias term ${3\sigma_{p}^{2} \over 2m}$ which is constant.
In other words, in this case one obtains a g.s. that resembles "stones" rather that quantum liquid drops as should be the case of nuclei.
Consequently, the approach insofar illustrated can describe only nuclear collisions for very short times or at very high excitation energy, well above the Fermi energy, when the Fermi motion plays a minor role in the dynamics.

\section{Introduction to the CoMD approach}

In the CoMD approach, we maintain the structure of a direct product for the N-body distribution but we change the equations of motion in this way:
\begin{eqnarray}
\dot{\langle{\bf r}\rangle}_{i}&=&\frac{\partial H}{\partial
\langle{\bf p}\rangle_{i}}\\ 
\dot{\langle{\bf p}\rangle}_{i}&=&- \frac{\partial H}{\partial
\langle{\bf r}\rangle_{i}}+{\bf C}_{i}+{\bf \lambda}_{i}+
{\bf P}_{i}+{\bf R}_{i}\\
\overline{f}_{i} & \leq & 1 
\end{eqnarray}
The meaning of the different terms ${\bf \lambda}$, ${\bf P}$ and ${\bf R}$ are explained in the following. 
They in general represent transformations of the centroid of the Gaussian nucleonic distribution function. 
To simplify the notation the angled brackets of these variables will be omitted in the subsequent notations.  

The set ${\bf P}_{i}$ represents impulsive forces, introduced for the first time in \cite{comd}, which perform a multi particle scattering for the generic particle $i$ whenever the related Pauli condition is violated. 
It is normally achieved through a series of two-body scatterings.
More precisely, the algorithm is the following: for each particle $i$ we define an ensemble of nearest particles $N_{i}$ identical to $i$ within the distance $3\sigma_{r}$ $3\sigma_{p}$ in the phase space.
If in a time step the average occupation number $\overline{f}_{i}>1$ (see(eq.2)) a loop is performed in
the algorithm over the particles belonging to the ensemble $N_{i}$ and, in the centre of mass of the generic couple $i,k$ of particles, the momenta are changed according to the following transformation:
\begin{eqnarray}
{\bf p_{i}}= -{\bf p_{k}} & \to & {\bf p'_{i}}=-{\bf p'_{k}}\\
|{\bf p_{i}}|=|{\bf p'_{i}}|  
\end{eqnarray}
The direction of the exchanged momentum $\Delta{\bf p}$ is chosen in a random way according to an isotropic distribution with respect to the initial direction ${\bf p_{i}}$. 
After this transformation the $\overline{f}_{i}$ is evaluated again, if it is less than or equal to one, the new coordinates  are accepted and the loop is stopped. 
If this is not the case, the coordinates of the involved particles are set to the old values and other attempts will be performed.
Normally 2 or at most 3 iterations are sufficient to satisfy the constraint on the atypical particle. 
Obviously, the algorithm is in a general loop running over all the particles and it can happen that one particle suffers multi-scattering process.
The procedure described keeps the dimensionality $N^{2}$ that is related to the two-body character of the particle interaction.
As can be easily deduced from the above relations, the transformations represent a series of elastic scattering processes that can mimic the repulsive character of the correlations associated to the anti-symmetric wave functions (exchange forces and vanishing of the N-body wave function at short distance in phase space). 

We observe that the Montecarlo procedure illustrated, which affects on average 10$\%$ of the total number of nucleons per time step, satisfies the Pauli requirement in a quite general way without introducing others correlations not directly related the Pauli principle itself. 

As a visual example of the action of the described procedure, in Fig.1  we sketch a situation in which 8 nucleons with same charge occupy close positions in phase space. 
The dashed circles represent the projection of a spherical volume $h^{3}$ in momentum space.
The arrows indicate the intrinsic spin. On the right, the occupation numbers related to some particles are also shown.
It is clear that the configuration shown in the upper panel is forbidden. 
In particular, the nucleon $1$ is very close to another one $a'$ with same spin. 
This kind of configuration appears spontaneously during the time evolution of the system if the ${\bf P}_{i}$ forces are inactive.

The bottom panel represents a possible configuration reached after the action of ${\bf P}$. 
The related forces perform a rotation (or scattering) in momentum space between the particles $1$ and $a'$ followed by a second rotation between $1$ and $2$.
The new configuration is now in agreement with the constraint on the occupation numbers. In particular it is equivalent to the previous one concerning the potential energy (we do not change their positions ${\bf r}_{i}$ and we use a momentum 
independent interaction) and the kinetic energy, therefore the total energy is exactly conserved.
Obviously the multi-particle transformation preserves the total momentum (action and reaction forces).

\section{Searching for "ground state configurations"} 

As happens for all problems that involve constraints, the depicted strategy can be successful in to obtaining the solution of the problem at each time step, if the initial conditions of the system satisfy the constraints.
The problem of the proper initial conditions involves the structure of nuclei in the "ground state"(g.s.) configurations.

For this purpose we make effective the small parameters ${\bf \lambda}_{i}$ ($|\lambda_{i}|\simeq 0.001$) which have the meaning of friction forces.  
We start with nucleons distributed in a hyper sphere of radius $R$ and $P_{F}$ (Fermi momentum), after which, we follow a localized procedure of  "cooling" and "warming" coupled with the constraint. At each time step, if for a particle $k$ $\overline{f_{k}}<1$, then for all the neighboring identical particles, $\lambda_{i}$ will be set to a negative value (cooling).
 On the contrary case ( $\overline{f_{k}}\ge 1$), $\lambda_{i}$ will be set positive. 
This procedure, applied for time intervals typically of some hundred of fm/c will, decrease the total energy to the minimum value. 
After this stage follows a "stabilization" phase (the $\lambda$ coefficients are set to zero). 
The microscopic configurations that are accepted as good replications  for the g.s. are those which are stable (no particle emission and nuclear radius stability within $10\%$) for a time interval of the order of some thousand of fm/c.
In ref.\cite{comd} the properties of the ground state of large class of nuclei with mass number ranging from N=30 to 197 have been studied to establish the model parameter for the two body interaction (including also the value of the Gaussian parameters $\sigma_{r}$=1.13fm, $\sigma_{p}=0.43\hbar$) in order to obtain an average binding energy of 8 MeV/nucleon and nuclear radii $R=1.2\cdot N^{1/3}$ (within 10$\%$) and with
an average compressibility of 200 MeV. 
The possibility to obtain stable configurations with the right macroscopic properties of nuclei it is directly linked with the properties of the effective interaction used and therefore with the dynamics related to the first term of eq.(2).
 
As an example of the results obtained for the g.s. configurations, we show in Fig.2, as open dots, the average occupation function $F(E_K)$ of the kinetic energy $E_K$ related to the centers of the Gaussian distribution for the $^{112}Sn$ nucleus.
These results are obtained after a "cooling", and "warming" procedure for a time interval of 650 fm/c. 
The average is computed on 100 configurations. 
The close connection to a Fermi distribution is clearly evident. Deviations from the typical stepwise shape, which characterizes the Fermi distribution at zero temperature, are related to the finite size effect of the investigated system and to the presence of N-body correlations arising from the 2-body short and long range (Coulomb) interaction. 
The error due to the finite number of configuration is of the order of $\pm 15\%$ at energies lower than 10 MeV. 
At energy higher than 30 MeV the errors are greater.
They are not shown in the picture to make the comparison clearer.
 
As previously observed, a simple "cooling procedure" without applying the coupling procedure with condition on the occupation numbers, obviously will produce a $\delta$ function centered at zero corresponding to a configuration in which the total effective kinetic energy related (first term in esq.(6))is zero.

In the same figure we show, as full dots, the $F(E_K)$ values related to the normal constrained dynamical evolution, without frictional forces, after 1500 fm/c. 
The distribution is very similar to the previous one and shows that the selected configurations can represent good g.s. samples. 
It demonstrates also that the algorithm is able to preserve the features of the Fermonic system during the time evolution. 

Finally, with crossed points we show, starting always from the same  configurations, the distribution obtained after 300 fm/c when the constraint on the occupation number $\overline{f}_{i}$ is suppressed.
It is clearly evident that in this case the Fermi-like behavior is not maintained, rather it appears to have a Boltzmann-like shape.
\vskip 5pt
\noindent
\section{CoMD II calculations in nucleus-nucleus collision}

In this section we present, as an example, some results obtained for $^{40}Ca+^{40}Ca$ collisions at 35 MeV/nucleon. 
In particular, by comparing the results obtained with experiment, we aim to emphasize the role of the constraint on the dynamics (see also ref. \cite{comd}). The nucleus-nucleus collision is simulated by taking into account the contribution related to an impact parameter window b from 0 to 8 fm. The initial configurations of the two $^{40}Ca$ nuclei are always
 good g.s. configurations obtained as described in the previous section.
In Fig.3 we show the charge distribution for the system under study in which the dynamical evolution is realized without the constraint. In the same figure the squared symbols represent the experimental results.
The calculations are plotted at t=3000 fm/c and t=300 fm/c. 
They show essentially the binary character of the reaction while the experimental data show a significant production of intermediate-mass fragments (IMF with Z$>$2). The binary behavior demonstrates a transparency effect in the calculations. The typical "U" shape is due to a large production of projectile- and target-like fragments formed in the first moments (bottom panel) of the interaction. After the formation, the two excited systems cool by emitting particles and the main bump in the charge distribution is displaced through lower charge values. 
This kind of "transparency" is caused by the lack of nucleon-nucleon collision (term C in eq.5) due to the overcrowding of the phase space (see also Fig.2, star symbols), which generate a strong suppression (about a factor 3) of the nucleon-nucleon cross section (Pauli blocking factor). 
In Fig.4 we display the same kind of comparison with calculations which include the constraint. The calculations are now in satisfactory agreement with the data. The increased rate of collisions produces a larger "stopping" and the related  compression-decompression effect. This effect, in turn, gives rise to mechanical instability generated the nucleon-nucleon field leading to the multi-fragmentation of the total system (IMF production)\cite{comd1,buu,bon}.

\section{CoMD II -Total angular momentum conservation}

In this section we will discuss the last term ${\bf R}$ added to the equations of motion (9). As mentioned in the introduction, the presence of the strong repulsive and short-range nuclear interaction, represented through the term ${\bf C}$, gives rise to another problem. Classical turning points make complications in the numerical integration of the equations of motion.
On the other hand, from a quantum point of view we have the lack of the trajectory concept. These circumstances could make it appropriate to simulate the effect of the nucleon-nucleon short-range repulsion (hard core) through elastic scattering process (and inelastic at the higher energies). In this case two nucleons, according to the related cross section, if close enough each other, can change their relative momentum, changing therefore the direction of motion.
In this way there is no correlation between the position of nucleons and the scattering angle and the quantum effect related to the finite spreading of the nucleon-nucleon wave function (branching) can be simulated. 
On the other hand, the loss of correlation generates a process that does not preserve the relative total angular momentum of the nucleons. 
Therefore the simulation globally violates the conservation law related to the total angular momentum ${\bf L}$ of the system (the angular momentum contribution related to the nucleon intrinsic spin is fixed) as happens in others N-body microscopic approach.  
This can be easily seen through the following relation valid for a scattering process in the centre of mass of the generic nucleon couple $1$ and $2$:

$$
{\bf L}_r^{in}=({\bf r}_{1}-{\bf r}_{2})\times{\bf p}_r^{in}\rightarrow
({\bf r}_{1}-{\bf r}_{2})\times{\bf p}_r^{fn}={\bf L}_r^{fn}
$$ 
and
$$
|{\bf p}_r^{fn}|=|{\bf p}_r^{in}| 
\;\; 
{\bf p}_r^{fn}\ne{\bf p}_r^{in}\rightarrow {\bf L}_r^{in}\ne{\bf L}_r^{fn}
$$
This problem can become prominent at high energy when the collision regime is important and, in general, in all the microscopic approaches, like in the CoMD model, that use random forces. 
!!*******In particular, in CoMD model, the problem is present also because we use multiple scattering processes to satisfy the 
!!Pauli prescriptions.

Before illustrating the transformation ${\bf R}$ used in the general case, we briefly discuss what we call a "trivial" solution to the problem.

The conservation of the angular momentum can be easily restored at a microscopic level if:
$$
{\bf p}_r^{fn}-{\bf p}_r^{in}={({\bf r}_{1}-{\bf r}_{2}) 
\over |({\bf r}_{1}-{\bf r}_{2})|}\Delta p
$$
This means that the change of the relative momenta in the scattering process is parallel to the relative distance between the nucleons.
This choice restores a one to one correspondence between the impact parameter b of the nucleon-nucleon collision and the scattering angle $\theta$.
In particular the related angular distribution for the scattering process is:
\begin{equation}
{d\sigma \over \d \Omega}=
{b \over \sin(\theta)}|{db \over d \theta}|=
{1 \over 4}R_{int}^{2}
\end{equation}
We have therefore an isotropic behavior.
 
We do not follow this solution because it appears quite restrictive for different reasons: the restoring of the one to one correspondence destroys the so-called "branching" effect; moreover, the behavior of the angular distribution for the scattering process is fixed. Therefore, in general it can deviate from the behavior suggested by the data related to the free nucleon-nucleon scattering or from more complicated calculations that take in to account in medium effects. 
In such cases, in fact the angular distribution can show marked anisotropy.
Finally, if we fix the relation between b and $\theta$ we do not have enough degree of freedom to satisfy the Pauli requirement through the multiple scattering processes given by the transformation ${\bf P}$.

These motivations therefore orientated us to choose a solving strategy that acts at a collective level. 
We search for a transformation ${\bf R}$ which involves the ensemble $C$ of nucleons which during a given time step have undergone a collision process.
In other words, if possible, we do not want to assign a particular role to few nucleons.
This ensemble of nucleons are characterized by the position of the center of mass (c.m.) in phase space with coordinates ${\bf r}_{C}$, ${\bf p}_{C}$.
In the c.m. reference system the total kinetic energy will be indicated by $T_{C}$. 
In the following all the nucleon coordinates belonging to the ensemble C are computed in this reference system.
The related algorithm can be summarized in the following steps:
\vskip 5pt
I) Step
\vskip 1pt 
We indicate with $\Delta\bf{L}$ the dissipated angular momentum in the considered time step and we search for a collective angular velocity vector ${\bf \omega}$ such that:
\begin{equation}
\Delta{\bf L}=I{\bf \omega}
\end{equation}
where $I$ represents the inertia tensor of the ensemble C.
With the collective angular velocity $\bf{\omega}$ we correct the momenta of the particles:
\begin{equation}
{\bf p^{'}}_ {k}={\bf p}_ {k}+m{\bf r}_ {k}\times{\bf \omega}
\:\:\:\:\:\:\:\:\:\:\:\:\:\:\:
k\subset C
\end{equation}
 
This first step restore the total angular momentum  conservation but it will change (just due to the rotational energy) the total energy of the system and in particular the value of $T_{C}$.
Therefore, in the next step, we have to change the momenta to restore energy conservation without changing the total angular momentum. Thence we perform only a scaling $\alpha$ of the radial momenta. Moreover the total momentum ${\bf p}_{C}$ has to remain unchanged.
\vskip 5pt
II) Step
\vskip 5pt
\begin{eqnarray}
{\bf p}^{''}_ {k} & = &{\bf p}^{'}_{k}+\alpha { ({\bf p}^{'}_{k}{\bf r}_ {k})
{\bf r}_{k} \over {r}^{2}_{k}} \\
{\bf p}^{'''}_ {k} & = & {\bf p}^{''}_{k}-\sum_{k}{
{\bf p}^{''}_ {k} \over {N}_{C}} \\  
 \sum_{k}{{\bf p}^{2'''}_{k} \over 2m}-T_{C} & = & \epsilon
\end{eqnarray}
with $k\subset C$.
The problem posed by the above 3 equations implies an iterative procedure depending on the scaling parameter $\alpha$ to minimize the absolute value of $\epsilon$.
In about 98$\%$ of the cases, this procedure solves the problem obtaining $min(|\epsilon|)=0$ within the computer accuracy.
In the other cases, starting from the previous step when the condition for $min(|\epsilon|)$ has been determined, we sort three particles of the ensemble C in such way to solve the algebraic system given by seven equations: 3 scalar equations for total spin, 3 for total momentum conservation and one for the total energy, without further scaling hypothesis.  

The above procedure does not work if the number of particles belonging to the ensemble C is less than 3. In this quite improbable case, we include in the ensemble C the necessary number of particles (which have not experienced a collision process in the considered time step) choosing them between the nearest particles to the c.m. position ${\bf r}_{C}$ of the ensemble C.

In the method illustrated, we have supposed that our system of N-particles is compact. That is the particles of C belong only to one cluster of nucleons.
This is not the case when, for example, we have the multi-break up of the hot compound system. In this case many clusters can be produced and, in a short time, they will be far apart.
If we include in the ensemble C all the colliding particles belonging to the different clusters, the procedure described will introduce unphysical long-range interactions.

Therefore in the common situation the transformation ${\bf R}$ has to be applied to all the ensembles C,C',C''...belonging to the different clusters. This means that in the CoMD II code the routine that defines the topology of the system (based on a coalescence model) has to be called at all time steps. 
This last request can be implemented by use of the modern parallel program architecture.

We conclude this section by comparing in Fig.5 the time dependence of the absolute value of the total angular momentum computed for one simulation of the collision $^{40}Ca+^{40}Ca$. 
The energy and the impact parameter are $E_{lab}=220 MeV$ and  $b=2 fm$ respectively. 
The dotted line represents the results obtained with the first version of the CoMD model while the continuous line is from the last version, CoMD II, implemented with the transformation ${\bf R}$.

\section{An illustrative example - $^{40}Ca+^{40}Ca$ fusion
 excitation function}

Practically in all the nuclear processes induced by heavy ion collisions, the role played by the angular momentum is crucial in to determine the main behavior of the reaction mechanism. On the other hand, also from a more static point of view, for a given temperature, the stability of the nuclear liquid drop depends on the amount of its spin, which strongly affects also the particle decay.
A typical example is the case of nuclear fission (or more generally the binary decay) which for most nuclei is hindered at zero temperature and zero spin but it can happen above some critical L value.  
Obviously, these effects can modify the fusion cross-section that represents, in some sense, the complementary process to the binary break-up of the compound system.

In Fig.6a we show the excitation function of the fusion cross section for the $^{40}Ca+^{40}Ca$ system.
From the figure it is possible to see, at the higher energy, about 30$\%$ reduction of the fusion cross section when
the $\bf{L}$ conservation law is taken into account.  
The fusion cross section has been evaluated by taking in to account the cross section for heavy residues production.
The crossed points represent data from ref.\cite{exp1} while the squared ones are data from ref.\cite{exp2}. 
This lowering is substantially due to about  $40-50\%$ increase of the binary process for the $\bf{L}$  conserved case (see Fig.6b)). 

We conclude this section by observing that in general in microscopic approaches a great effort is made in trying to obtain the right predictions for the fusion cross section. 
These efforts involve, generally, a search for the best parameters of the effective interaction (especially the ones that describe the surface term) able to produce the measured value. Work is in progress in fact to reproduce the point at  $E_{cm}=150 MeV $ even if, apart from this point, we note for the investigated system the lack of others fusion cross section measurements in this energy region.

Nevertheless, the results shown in this section highlight how it is quite important to work in a scheme in which the total angular momentum is conserved before reaching any kind of conclusion about the parameters that better describe the effective nucleon-nucleon interaction.

\section{Conclusive Remarks}
In this paper we have shown a recent development of the constraint dynamics, introduced in \cite{comd}, in modeling the N-body approach in nuclear physics.
The Constrained Molecular Dynamics model is based on the use of impulsive forces that constrain the semi-classical dynamics to satisfy the Pauli principle and, in CoMD II, also the total angular momentum conservation. Satisfaction of this quit 
fundamental law has been undertaken in other semi-classical microscopic approaches in which the hard-core repulsive interaction is simulated through nucleon-nucleon collision processes. 
The choice of this strategy corresponds to the usage of transformations on the nucleon coordinates in momentum space which allow to work with an  $N^{2}$ dimensionality, imposed by the two-body character of the used effective interaction, rather that a $N^{4}$ dimensionality as prescribed  by an antisymmetric dynamics. 
This choice has the obvious advantage to shorten  drastically the computing time, keeping the essential features related to the fermionic dynamics.
Finally, as an example,  we have shown the pronounced differences in the fusion cross section excitation function computed with the two CoMD versions due to the restoration of the total angular momentum conservation law in CoMD II.


\vfill
\eject
\noindent

Fig.1. - Typical configuration for 8 nucleons with same charge 
in momentum space before 
(upper panel) and after (bottom panel) the action of the transformation 
${\bf P}$. The dashed circles represent the projection of a spherical
volume $h^{3}$ in momentum space. On the right the values of the 
occupation numbers
$\overline{f}$ for different nucleons are also reported.
\vskip 20pt
\noindent
Fig.2. - With empty circles we show as function of the nucleon kinetic energy the average occupation function $f(E_{k})$ for the obtained "ground state" configurations of the $^{112}Sn$ nucleus. The others symbols are referred to the obtained distribution after a time interval $\Delta t$ of dynamical evolution with and without the action of the constraint(see the legend).
\vskip 20pt
\noindent
Fig.3. –Comparison \cite{comd} between the experimental isotope distribution measured for $^{40}Ca+^{40}Ca$ system at 35 MeV/nucleon \cite{caex} and theoretical prediction performed according to our approach without the constraint on the Pauli prescription during the dynamical evolution.
\vskip 20pt
\noindent
Fig.4. –- Same as Fig.4 but for the CoMD model \cite{comd}. 
\vskip 20pt
\noindent
Fig.5. - Absolute value of the total angular momentum as function of time for one realizations of the $^{40}Ca+^{40}Ca$ system at $E_{cm}=110 MeV$ and impact parameter $b=2 fm$. In the figure we
show the results of the model calculations with the use of the ${\bf R}$ transformation (CoMD II) (continuous line) and without (dashed line).
\vskip 20pt
\noindent
Fig.6. - a) Fusion cross section as function of the c.m. incident energy for the $^{40}Ca+^{40}Ca$ system. The filled circles indicate the results obtained with CoMD II while the empty circles represent the results obtained without the transformation ${\bf R}$. The crossed and squared points represent experimental data taken from ref.\cite{exp1} and \cite{exp2} respectively.

-b) at $E_{cm}$ 130 MeV we show 
the probability distribution as function of the impact parameter b for dissipative binary reactions in the above two cases.

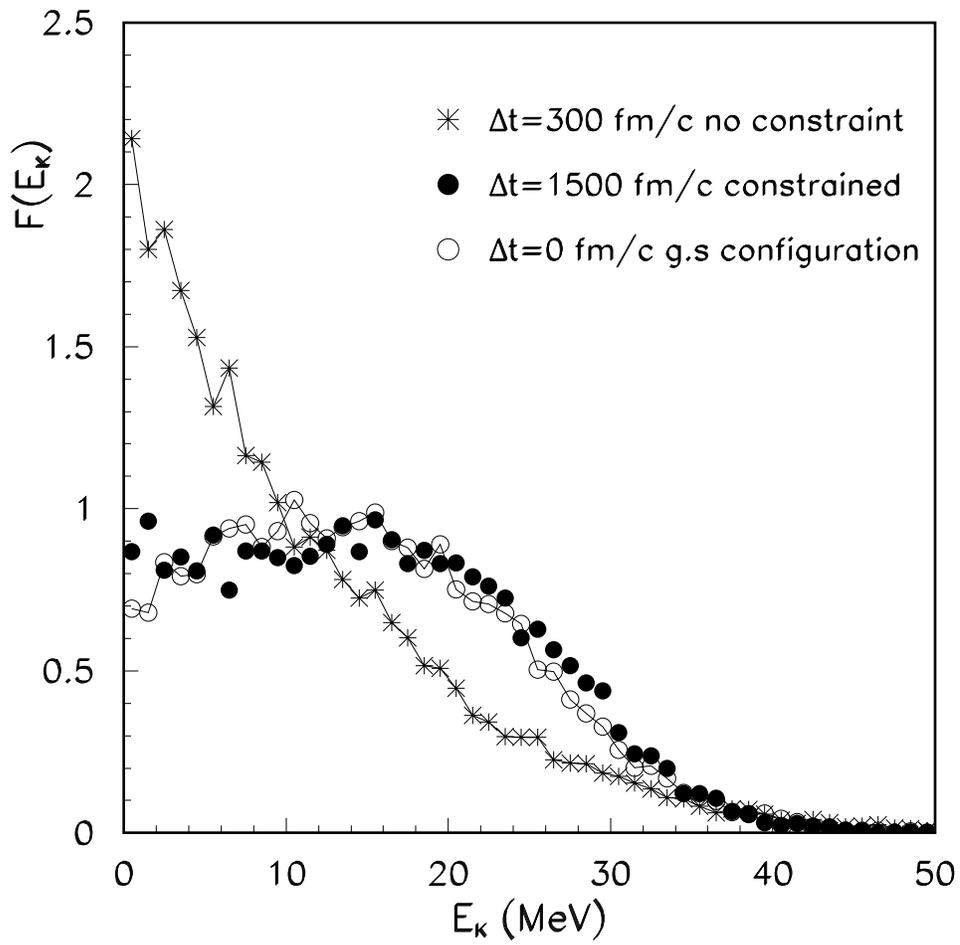
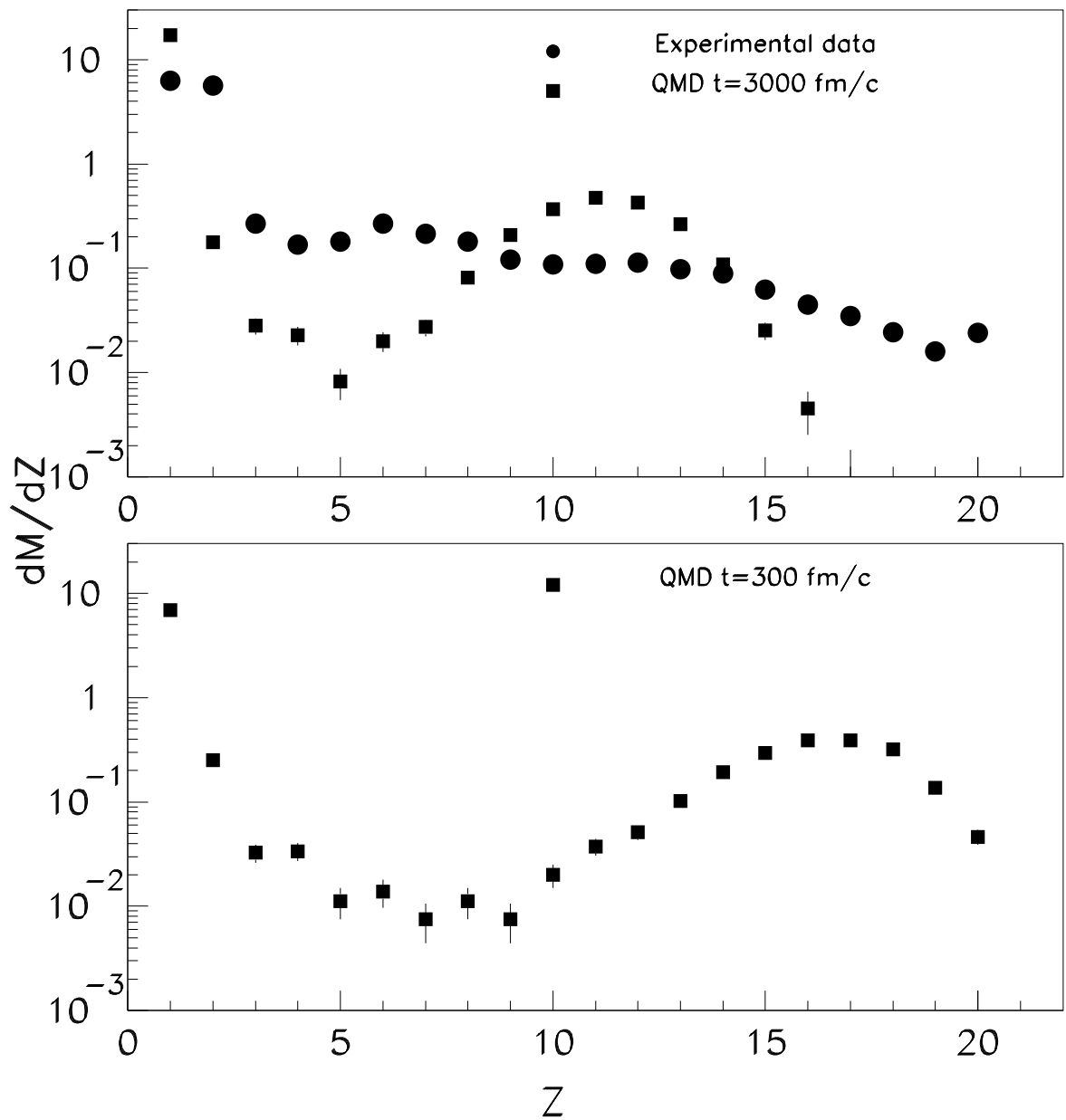
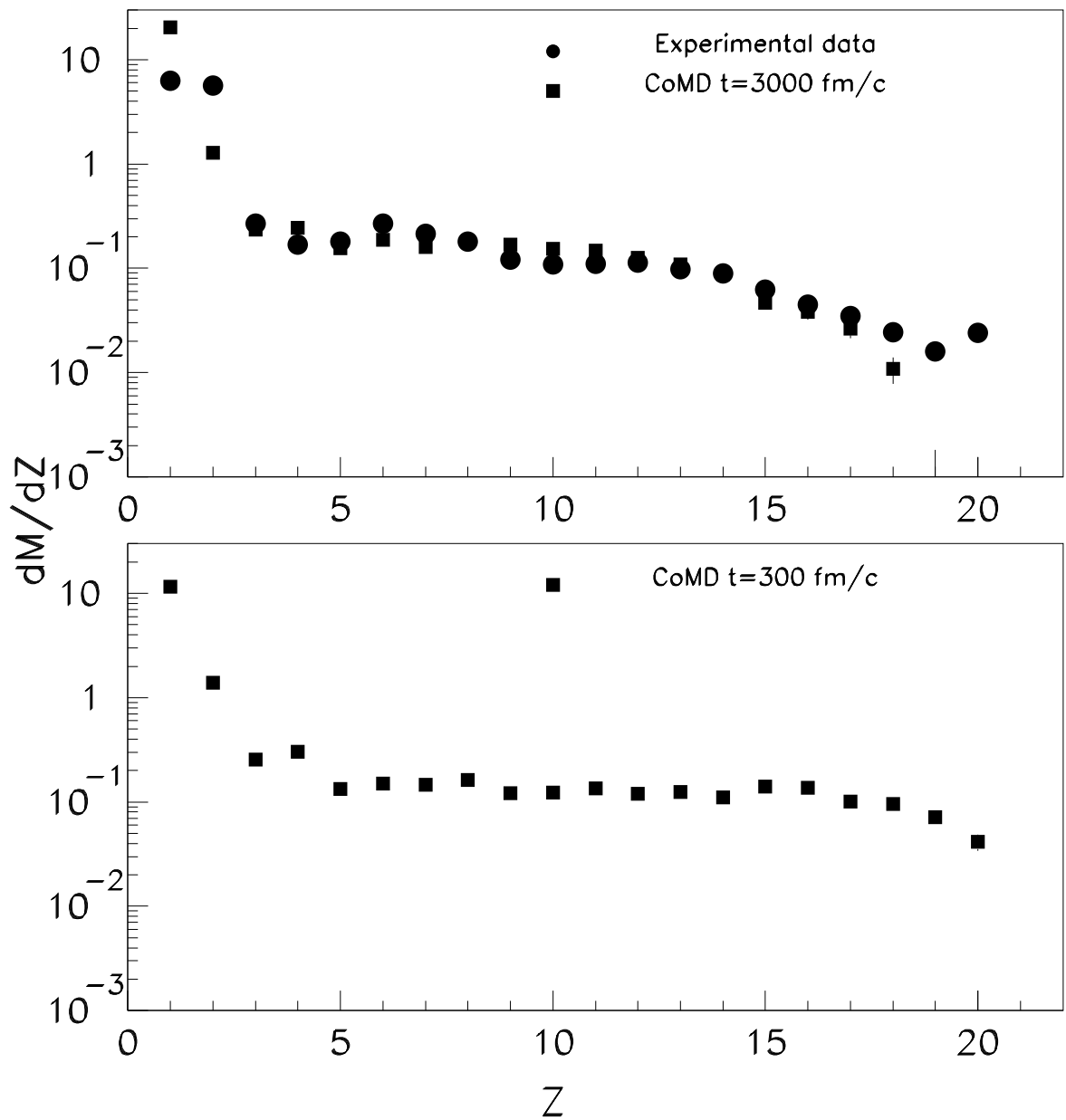
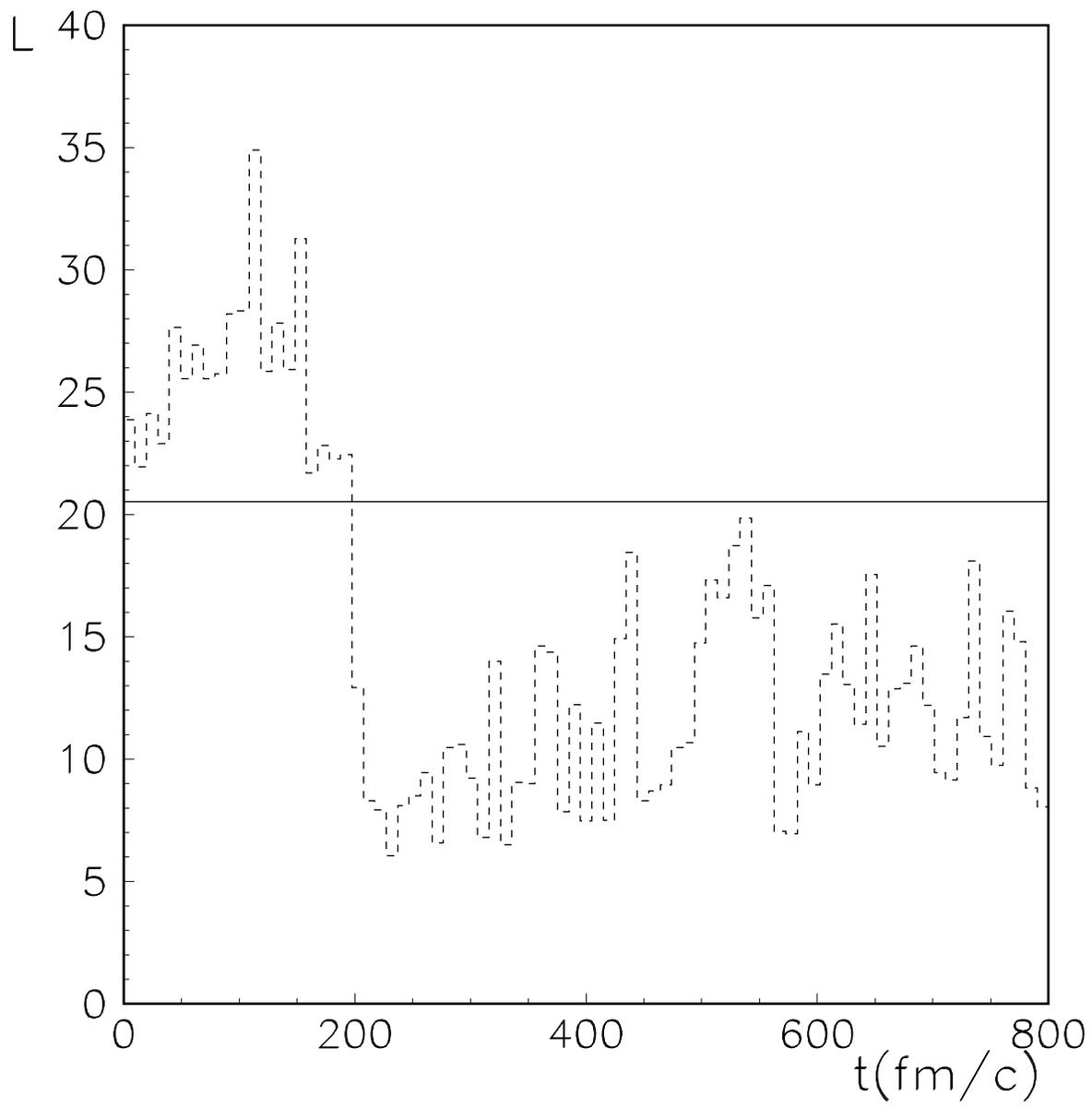
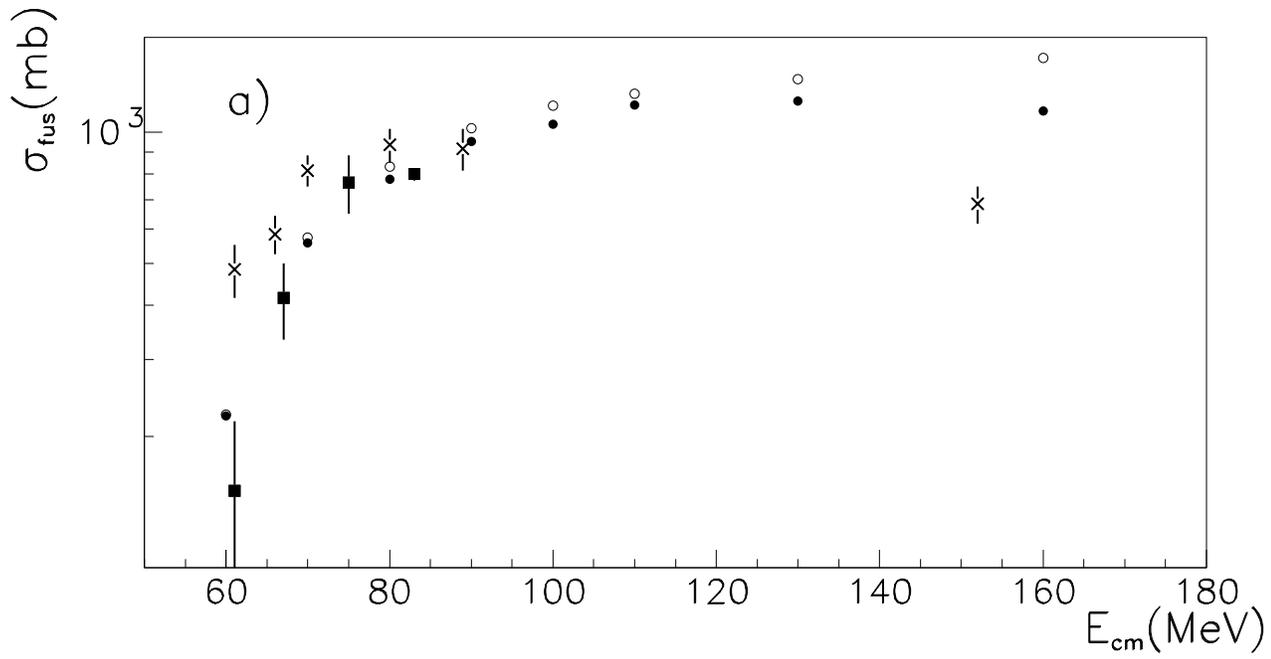
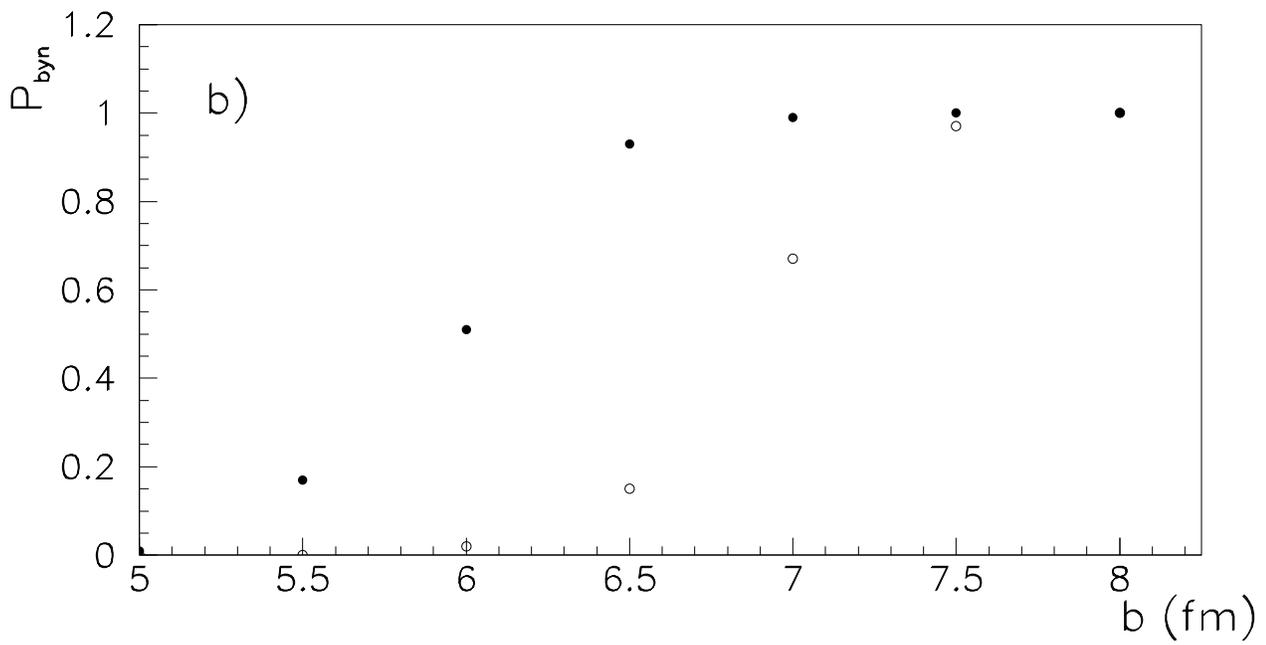
\end{document}